\documentclass[%
 reprint,
 amsmath,amssymb,
 aps,
]{revtex4-2}

\usepackage{graphicx}
\usepackage{dcolumn}
\usepackage{bm}

\usepackage{multirow}
\usepackage{url}
\usepackage{color}
\usepackage[misc]{ifsym}

\begin{document}

\preprint{APS/123-QED}

\title{Quasiparticle band alignment and stacking-independent exciton in MA$_2$Z$_4$ (M = Mo, W, Ti; A= Si, Ge; Z = N, P, As)}

\author{Hongxia Zhong,\textsuperscript{1*} Guangyong Zhang,\textsuperscript{1} Cheng Lu,\textsuperscript{1} Shiyuan Gao\textsuperscript{2*}}
\affiliation{\textsuperscript{1}School of Mathematics and Physics, China University of Geosciences, Wuhan 430074, China}

\affiliation{\textsuperscript{2}Institute for Quantum Matter and Department of Physics and Astronomy, Johns Hopkins University, Baltimore, Maryland 21218, USA}%

\email{zhonghongxia@cug.edu.cn}
\email{sgao45@jh.edu}



\begin{abstract}
Motivated by the recently synthesized two-dimensional semiconducting MoSi$_2$N$_4$, we systematically investigate the quasiparticle band alignment and exciton in monolayer MA$_2$Z$_4$ (M = Mo, W, Ti; A= Si, Ge; Z = N, P, As) using ab initio GW and Bethe-Salpeter equation calculations. Compared with the results from density functional theory (DFT), our GW calculations reveal substantially larger band gaps and different absolute quasiparticle energies, but predict the same types of band alignments. Using MoSi$_2$N$_4$/WSi$_2$N$_4$ heterostructure as a model, we find that the quasiparticle band energies are insensitive to environmental screening, with band edge energies changing about 0.15 eV, leading to a  $\sim $ 4$\%$ reduction in the quasiparticle band gap compared to free-standing monolayers. Finally, inspired by recent interests in moiré excitons, we study quasiparticle and exciton properties of MoSi$_2$N$_4$/WSi$_2$N$_4$  bilayers with different stacking configurations. The optical dipole oscillator strength and energy of intralayer excitons are almost independent of the stacking configuration, while the interlayer exciton is dark due to negligible electron-hole overlap. The quasiparticle band alignment and stacking-independent exciton obtained in this work are important for designing heterojunction and high-efficiency optoelectronic devices based on MA$_2$Z$_4$.
\end{abstract}

\maketitle


\section{Introduction}

Owing to the depressed screening and strong electron-electron interactions, many-electron effects are more pronounced in two-dimensional (2D) materials compared with their bulk phases\cite{ugeda2014giant,mak2010atomically,tran2014layer,qiu2013optical}. The enhanced many-electron effects are expected to increase the quasiparticle band gap and cause electron-hole pairs to form strongly bound excitons\cite{spataru2004excitonic,zhong2015quasiparticle}. Indeed, for most 2D semiconductors, the quasiparticle band gaps are larger than the single particle band gaps by 0.5-2 eV, and the exciton binding energies are at least an order magnitude larger than those of bulk structures\cite{choi2015linear,spataru2004excitonic,zhong2015quasiparticle}. Besides the precise band gap, the relative band edge energies between different semiconductors and corresponding band alignments are desirable for understanding fundamental physics and designing heterojunction devices\cite{liang2013quasiparticle,zheng2018band}.

During the process of 2D material growth, it needs a substrate, and the environmental screening from the substrate will significantly renormalize the quasiparticle band gap and exciton binding energy of the atomically thin system\cite{ugeda2014giant,qiu2017environmental,bradley2015probing}. For example, when black phosphorus is encapsulated between sapphire and $h$-BN, the substrate screening of sapphire reduces the binding energy of monolayer black phosphorus by as much as 70$\%$ and completely eliminates the presence of bound excitons in four-layer black phosphorus\cite{qiu2017environmental,li2017direct,zhang2017infrared,zhang2014extraordinary}. By stacking two monolayers together, van der Waals heterostructures are formed, inheriting and extending the physical properties of the original monolayers\cite{novoselov20162d,kunstmann2018momentum,geim2013van}. Each layer of the heterostructure has different orientations, and the relative orientation between two layers of a 2D crystal introduces a longer-period variation of local atomic arrangement, which is known as the moiré pattern\cite{tran2019evidence,alexeev2019resonantly}. It has been demonstrated that the transition-metal dichalcogenide (TMDC) heterobilayer moiré pattern can modulate the electronic band structure and induce multiple interlayer exciton resonances in experiments\cite{kunstmann2018momentum,tran2019evidence,choi2021twist}. Therefore, the effect of environmental screening and moiré pattern on quasiparticle band alignment and excitons are of fundamental interest in 2D semiconductors.

Here, we focus on the quasiparticle band alignment and exciton properties of the recently synthesized semiconducting MoSi$_2$N$_4$ and its family\cite{hong2020chemical,wang2021intercalated}, which is a member of another new family of 2D molybdenum nitrides without bulk phases. Monolayer MoSi$_2$N$_4$ shows an optical gap of $\sim $1.94 eV, and high intrinsic electron carrier mobilities (1200 cm$^2$/(v$\cdot$s))\cite{hong2020chemical}. The excellent mechanical properties and ambient stability of monolayer MoSi$_2$N$_4$ have inspired much follow-up works, and a number of interesting properties such as spin-valley coupling\cite{yang2021valley,li2020valley,cui2021spin}, high thermal conductivity\cite{mortazavi2021exceptional}, and piezoelectricity\cite{yu2021high} have been predicted. However, the quasiparticle band alignment of monolayer MA$_2$Z$_4$ has yet to be studied, and the effect of environmental screening and moiré pattern on quasiparticle band alignment and excitons is unknown.

In this work, we will fill in these blanks theoretically by first-principles GW-Bethe Salpeter equation (BSE) approach. Based on the electronic structure details, we discuss the absolute quasiparticle band alignment of monolayer MA$_2$Z$_4$ (M = Mo, W, Ti; A= Si, Ge; Z = N, P, As) and compare it with the result from density functional theory (DFT). It is found that the band-gap-center model is valid for the studied monolayer MA$_2$Z$_4$, leading to the same types of band alignments predicted by both GW and DFT calculations. We then study the environmental screening on the band alignment using the model MoSi$_2$N$_4$/WSi$_2$N$_4$ heterostructure, and find the quasiparticle band gap renormalization is very small, suggesting the weak environmental screening effect on the quasiparticle band alignments of MA$_2$Z$_4$. Finally, we focus on the excitons in MoSi$_2$N$_4$/WSi$_2$N$_4$ heterostructure with three local identified stackings to discuss the effect of moiré pattern. The optical dipole oscillator strength and energy of intralayer excitons are almost independent of the stacking configuration, while the interlayer exciton is dark owing to negligible electron-hole overlap.

The remainder of this paper is organized as follows: In Sec. II, we introduce the atomic structures of monolayer MA$_2$Z$_4$, and our computational approaches. In Sec. III, the quasiparticle band gap and electronic structure details are presented. In Sec. IV, the quasiparticle band alignment is given. In Sec. V, we discuss the environmental screening effect on the quasiparticle band alignment of monolayer MA$_2$Z$_4$. In Sec. VI, we focus on the low-energy excitons in MoSi$_2$N$_4$/WSi$_2$N$_4$ heterostructure with different stackings. Finally, the conclusion is summarized in Sec. VII.

\section{Atomic Structure and Calculation Method}

Compared with hexagonal TMDC, monolayer MA$_2$Z$_4$ also belongs to the hexagonal lattice but with lower symmetry (space group P$\bar{6}$m2 (No. 187)), as shown in Fig. 1. Along the z direction, it has a septuple atomic layers of Z-A-Z-M-Z-A-Z, which can be viewed as a MZ$_2$ layer passivated by two A-Z bilayers, leading to the excellent ambient stability of MA$_2$Z$_4$ monolayers\cite{hong2020chemical}. We fully relax monolayer MA$_2$Z$_4$ (M = Mo, W, Ti; A= Si, Ge; Z = N, P, As) according to the force and stress calculated by density functional theory (DFT) with the Perdew, Burke, and Ernzerhof (PBE) functional\cite{perdew1996generalized}, using the QUANTUM ESPRESSO package\cite{giannozzi2009quantum}. The van der Waals corrections (vdW-DF) are utilized to include the interlayer interaction in the constructed heterostructures\cite{otero2012van}. The ground state wave functions and eigenvalues are obtained from the DFT/PBE with norm-conserving pseudopotentials which include the semi-core states of transition metal M\cite{troullier1991efficient}. The plane-wave basis is set with a cutoff energy of 80 Ry with a 16 $\times$ 16 $\times$ 1 $k$-point grid. A vacuum space between neighboring layers is set to be more than 25 Å to avoid interactions between layers. Based on these parameters, the relaxed lattice constant for monolayer MoSi$_2$N$_4$ (2.910 Å) is in good agreement with previous calculations\cite{hong2020chemical,wang2021electronic}. The optimized lattice constants of MA$_2$Z$_4$ are summarized in Table I. One can see that the lattice constant increases as the radius of A or Z atoms increases. For example, the lattice constant increases from 2.910 Å in MoSi$_2$N$_4$ to 3.469 Å in MoSi$_2$P$_4$ to 3.617 Å in MoSi$_2$As$_4$, in agreement with increasing trend of the atomic radius of Z atoms. And the lattice constant of MoGe$_2$N$_4$ (3.035 Å) is slightly larger than that (2.910 Å) of MoSi$_2$N$_4$, owing to the slightly larger atomic radius of Ge than Si atom. On the contrary, the lattice constant is nearly not affected by the M atom, which is always around 2.920 Å for all MSi$_2$N$_4$ (M = Mo, W, Ti) monolayer. Therefore, primitive MSi$_2$N$_4$ monolayer can be adjusted to other MSi$_2$N$_4$ unit cell, forming heterostructures with negligible lattice mismatch.

The excited-state properties of the heterostructure are calculated by the GW approximation within the general plasmon pole model\cite{hybertsen1986electron}, which is reliable in obtaining the excitonic properties of 2D materials. The unoccupied conduction band number involved in calculating the dielectric function, self-energy, and absolute band edge is set to be 2000 after converge test. In solving the BSE, we use a finer $k$-point grid of 48 × 48 × 1 for converged excitonic states\cite{rohlfing2000electron}. All the GW-BSE calculations are performed with the BerkeleyGW code\cite{deslippe2012berkeleygw} including the slab Coulomb truncation scheme to mimic interactions between structures\cite{ismail2006truncation,rozzi2006exact}. For optical absorption spectra, only the incident light polarized parallel with the plane is considered due to the depolarization effect\cite{spataru2004quasiparticle,yang2007enhanced}.

\section{Electronic Structures}

\begin{figure}[h]
\centering
\includegraphics[scale=0.32]{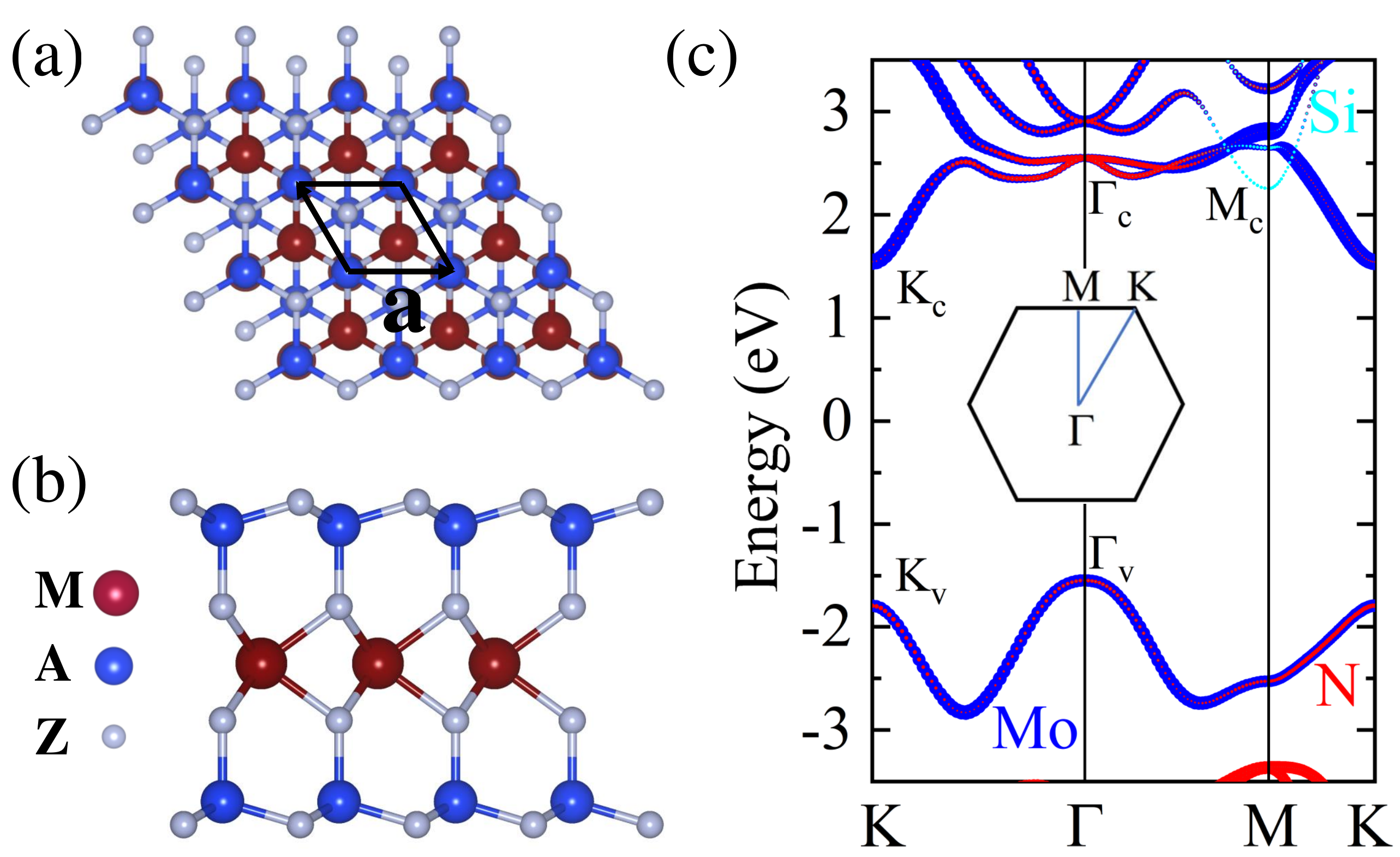}
\caption{Top (a) and side (b) views of atomic structure of monolayer MA$_2$Z$_4$. The primitive cell and lattice vector $\bm{a}$ are labeled in (a). Quasiparticle band structure (c) of monolayer MoSi$_2$N$_4$. The band gap center is set to be zero. Two highest valence band states K$_v$ and $\Gamma_v$, and three lowest conduction band states K$_c$, $\Gamma_c$, and M$_c$ are labeled. Here, we define the energy difference $\Gamma_v$ - K$_v$ as $\Delta_v$, and the energy difference $\Gamma_c$ – K$_c$ (M$_c$ – K$_c$) as $\Delta_{c1}$ ($\Delta_{c2}$).}
\end{figure}

We plot the quasiparticle band structure of monolayer MoSi$_2$N$_4$ in Fig. 1c as an example to discuss the general electronic features of studied MA$_2$Z$_4$ (M = Mo, W, Ti; A= Si, Ge; Z = N, P, As). It is shown that monolayer MoSi$_2$N$_4$ is an indirect bandgap semiconductor, with valence band maximum (VBM) at the $\Gamma $ point and conduction band minimum (CBM) at the K point. There is another local maximum of the valence band at the K point, whose energy is very close to the VBM at the $\Gamma $ point. Actually, this local maximum at the K point can increase to become the VBM in monolayer MoSi$_2$P$_4$ and MoSi$_2$As$_4$, and thus, the overall band structure turns out to possess a direct band gap\cite{hong2020chemical}, which displays much stronger photoluminescence accordingly. In order to analyze this subtle but important change in band structure, we define quantitatively the energy difference between the two highest valence band states $\Gamma_v$ - K$_v$ as $\Delta_v$ (see Fig. 1c), and summarize its values in Table I. 

\begin{table*}
    \centering
    \caption{Lattice constant $\bm{a}$, band gap $E_g$, energy difference $\Delta_v$, $\Delta_{c1}$, and $\Delta_{c2}$ (see Fig. 1c) for monolayer MA$_2$Z$_4$ (M = Mo, W, Ti; A = Si, Ge; Z = N, P, As).}
    \begin{tabular}{ c c c c c c c c c c }
    \hline
    \hline
    &&&&&&&&&  \\
    & a (Å)  & $E_g^{DFT}$ (eV)  & $E_g^{GW}$ (eV)& $\Delta_v^{DFT}$ (eV) &  $\Delta_v^{GW}$ (eV) &  $\Delta_{c1}^{DFT}$ (eV) & $\Delta_{c1}^{GW}$ (eV)   & $\Delta_{c2}^{DFT}$ (eV) & $\Delta_{c2}^{GW}$ (eV)  \\
    &&&&&&&&&  \\
    \hline
    MoSi$_2$N$_4$ & 2.910 & 1.796 & 3.090 & 0.309 & 0.233 & 1.029 & 1.074 & 0.681 & 0.296 \\
    
    MoSi$_2$P$_4$ & 3.469 & 0.701 & 1.324 & -0.251 & -0.605 & 1.919 & 1.418 & 0.705 & 0.054 \\
    
    MoSi$_2$As$_4$ & 3.617 & 0.612 & 1.126 & -0.199 & -0.522 & 1.846 & 1.244 & 0.710 & 0.684 \\
    
    MoGe$_2$N$_4$ & 3.035 & 0.927 & 2.052 & 0.601 & 0.547 & 1.001 & 0.739 & 1.404 & 1.672 \\
    
    WSi$_2$N$_4$ & 2.912 & 2.079 & 3.018 & 0.278 & 0.207 & 0.813 & 0.958 & 0.172 & -0.212 \\
    
    TiSi$_2$N$_4$ & 2.932 & 2.079 & 3.261 & 0.336 & 1.02 & 0.442 & 0.505 & -0.726 & -1.081 \\
    \hline
    \hline
    \end{tabular}
    \label{tab:my_label}
\end{table*}

We first discuss the energy difference at the DFT level. The positive value of $\Delta_v$ means the indirect band gap, and the negative value in monolayer MoSi$_2$P$_4$ and MoSi$_2$As$_4$ means the direct band gap. For most MA$_2$Z$_4$, the $\Delta_v$ value is positive and larger than 200 meV, which can be comparable to those in monolayer TMDC. Meanwhile, we have marked the energy difference $\Delta_{c1}$ ($\Delta_{c2}$) between the lowest conduction band states $\Gamma_c$ – K$_c$ (M$_c$ – K$_c$), and listed their values in Table I. Compared with the $\Delta_v$ values, the values of $\Delta_{c1}$ and $\Delta_{c2}$ are much larger, ranging from 0.681 to 1.919 eV, except for the $\Delta_{c2}$ in monolayer WSi$_2$N$_4$. The large energy difference is attributed to the strong hybridization of lowest conduction band. The K$_c$ state is dominated by M atoms, and $\Gamma_c$ state is mainly contributed by Z atoms, while M$_c$ state only contains A atoms (see Fig. 1c). It is the hybridized lowest conduction band that results in the dispersive CBM\cite{hong2020chemical}, which is very different from the balanced carrier mobility in TMDC.

Owing to the depressed screening and stronger electron-electron (e-e) interactions\cite{ugeda2014giant,mak2010atomically,tran2014layer,qiu2013optical}, significant self-energy enhancements are expected to be observed in monolayer MA$_2$Z$_4$. We thus apply the single-shot G$_0$W$_0$ approach to calculate the quasiparticle energy of those six monolayer MA$_2$Z$_4$ (M = Mo, W, Ti; A= Si, Ge; Z = N, P, As), and summarize the results in Table I. First, our calculated G$_0$W$_0$ band gap (3.090 eV) of monolayer MoSi$_2$N$_4$ is in good agreement with the value (3.190 eV) in previous studies\cite{liang2022highly}. Second, the GW correction significantly enlarges the bandgap for all studied MA$_2$Z$_4$. At the G$_0$W$_0$ level, the quasiparticle band gap of monolayer MA$_2$Z$_4$ is increased by 0.5-1.3 eV. Finally, the G$_0$W$_0$ band gap is significantly larger than HSE value (2.297 eV), while the measured A excitonic transition is around 2.3 eV\cite{hong2020chemical}. Compared with the invalid HSE calculation, the GW band gap may explain (see Section VI) the experimental photoluminescence spectra, verifying the important many body effects in monolayer MA$_2$Z$_4$.

Beside the band gap, we also discuss the effect of self-energy enhancements on the energy difference in Table I. Overall, the quasiparticle correction does not modify the sign of $\Delta_v$, due to the similar quasiparticle correction of M $d_{z^2}$ states for the whole highest valence band in monolayer MA$_2$Z$_4$. In details, for direct semiconductors ($\Delta_v$ $<$ 0), the $\Delta_v$ value is enhanced by the quasiparticle correction. The quasiparticle energy difference $\Delta_v$ of monolayer MoSi$_2$P$_4$ (MoSi$_2$As$_4$) increases from -0.251 (-0.199) eV at DFT level to -0.605 (-0.522) eV by GW calculation. By contrast, the $\Delta_v$ value for indirect semiconductors ($\Delta_v$ $>$ 0) is not significantly modified by the quasiparticle corrections. While for the lowest conduction band, the $\Delta_{c1}$ and $\Delta_{c2}$ are reduced by the quasiparticle corrections. These different self-energy corrections are caused by the nature of the involved electronic states K$_c$, $\Gamma_c$, and M$_c$. Generally, the localized electronic states enhance the overlap between wave functions and screened Coulomb interactions, resulting in a large self-energy\cite{zhong2015quasiparticle}. Therefore, the quasiparticle energy of the K$_c$ state (dominated by M $d_{z^2}$ states) is enhanced more, and the energy level is heightened more, leading to the reduction in $\Delta_{c1}$ and $\Delta_{c2}$ after quasiparticle correction.

\section{Quasiparticle band alignment}

It is well known that relative band edge energies between different semiconductors and corresponding band offsets are of fundamental interest in solid state physics and are indispensable for the design of heterojunction devices\cite{liang2013quasiparticle,zheng2018band}. We thus calculate the absolute band edge energy relative to the vacuum level. The absolute band edge energy at the DFT level is aligned with the vacuum level, which is set to zero. With the inclusion of self-energy corrections to the DFT eigenvalues, we obtain the absolute quasiparticle energy relative to the vacuum level. Because the convergence of the absolute quasiparticle band edge energy is slower than the band gap convergence, the GW calculations on the absolute quasiparticle band edge energy have to be checked carefully on the $k$ grid and empty bands to get reliable results. We increase $k$-point grid to 16 × 16 × 1 for relaxation and 48 × 48 × 1 for static calculations, and empty bands to 2000, and find that the absolute quasiparticle band edge energy changes slightly ($\sim $0.1 eV). Therefore, we calculate the quasiparticle band edge energy based on the above converge parameters.

\begin{figure}[h]
\centering
\includegraphics[scale=0.36]{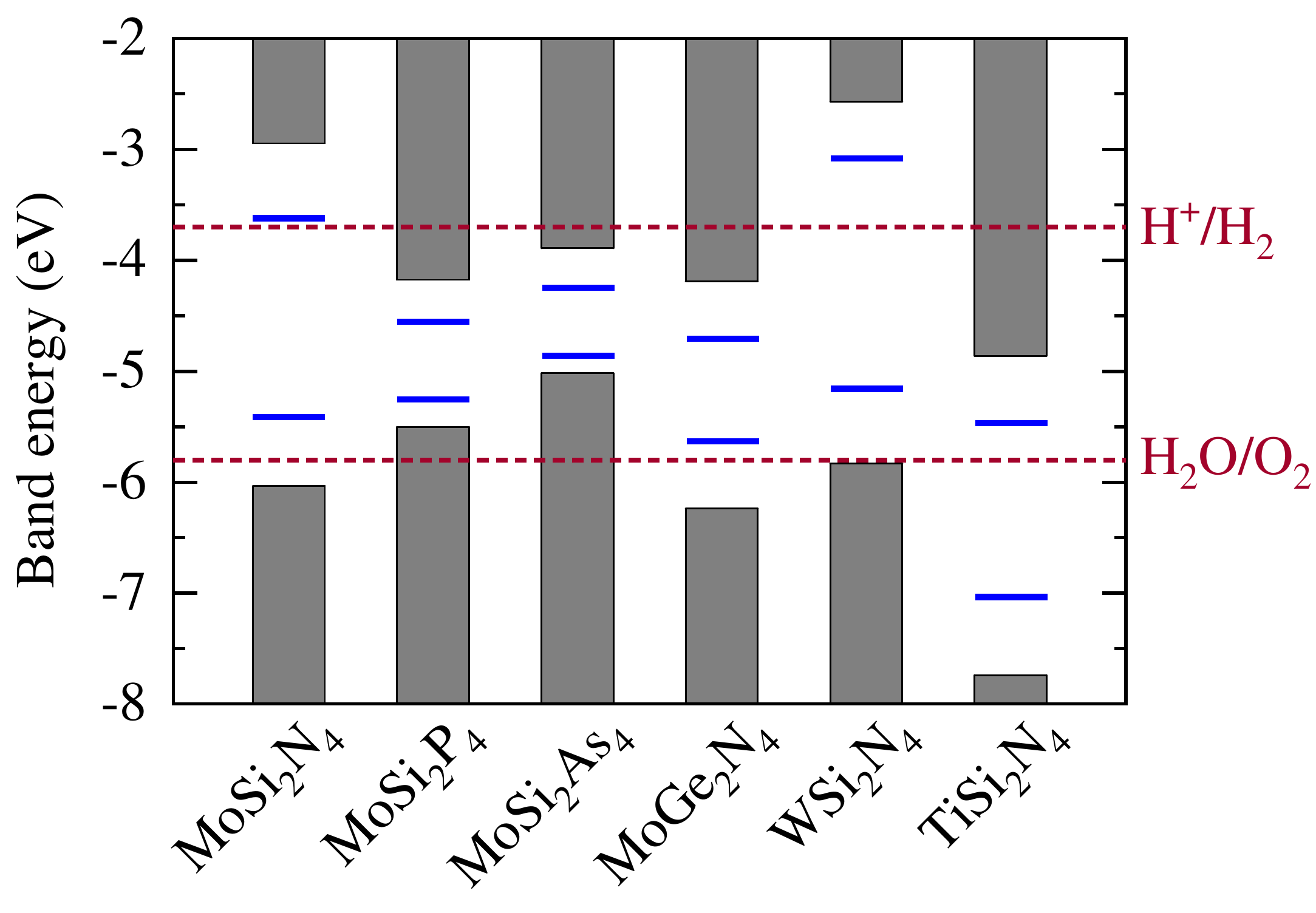}
\caption{The absolute band edge energies of calculated monolayer MA$_2$Z$_4$ relative to the vacuum level. The grey-shadow regions (blue dashed lines) stand for the fully converged GW (DFT) results. The water reduction (H$^+$/H$_2$) and oxidation (H$_2$O/O$_2$) are marked by the red dashed lines, respectively.}
\end{figure}

Figure 2 presents the absolute quasiparticle band edge energies of monolayer MA$_2$Z$_4$ (M = Mo, W, Ti; A= Si, Ge; Z = N, P, As), in which the DFT results are also listed for reference. First, the values of GW calculated band offsets are larger than those from DFT owing to the large quasiparticle band gap correction. Second, the general trend of the evolution of the band edge energies is similar for both DFT and GW results. For instance, the VBM of monolayer MA$_2$Z$_4$ gradually increases as Z varies from N to P to As, or M varies from Mo to W. As a result, the qualitative types of band alignments for these MA$_2$Z$_4$ from DFT have not changed. Both DFT and GW calculations predict that the MoSi$_2$N$_4$/WSi$_2$N$_4$ (MoSi$_2$N$_4$/MoSi$_2$As$_4$) heterostructure has a type-II (type-I) band alignment. Third, monolayer MoSi$_2$N$_4$ and WSi$_2$N$_4$ are suitable for water splitting, as a result of lower VBM (higher CBM) than the oxidation energy of -5.8 eV (reduction energy of -3.7 eV)\cite{ullah2018electronic}. Finally, both MoSi$_2$N$_4$ and WSi$_2$N$_4$ are hardly doped n-type owing to their quasiparticle CBMs above the pining energy of -4.0 eV, but easily doped p-type because of the suitable quasiparticle VBMs above the other pining energy of -6.0 eV\cite{hu2020p}. We thus can conclude that for 2D heterojunctions of our studied 2D MA$_2$Z$_4$ with other semiconductors, the DFT calculation is enough to assess the type of band alignment, and the GW method is necessary to obtain the quantitative band offset or other related properties.

In order to avoid the costly GW calculation on the quantitative band offset, the band-gap-center approximation has been proposed to estimate the absolute quasiparticle band edge energy with the assumption that the self-energy correction shifts both VBM and CBM in inverse direction with similar amounts, such as in monolayer TMDCs\cite{liang2013quasiparticle,zhong2016interfacial}. We find that this model is also valid for our studied monolayer MA$_2$Z$_4$ (M = Mo, W, Ti; A = Si, Ge; Z = N, P, As). The band-gap-center approximation gives nearly the same band edge energy as the costly direct GW calculation in Fig. 2. Taking monolayer MoSi$_2$N$_4$ as an example, the VBM (CBM) from band-gap-center model is -6.062 (-2.972) eV, in line with the corresponding value of -6.034 (-2.944) eV from direct GW calculation. On the other hand, one must be cautious when applying the band-gap-center approximation to other MA$_2$Z$_4$. For example, for monolayer CrSi$_2$N$_4$, the VBM at around -5.4 eV is almost unchanged by the G$_0$W$_0$ self-energy, while the CBM is shifted up by 1.113 eV with the quasiparticle correction. The unchanged VBM may be related to the much larger screened exchange and coulomb hole contribution in CrSi$_2$N$_4$ compared with those in MoA$_2$Z$_4$ and WA$_2$Z$_4$. We note that this asymmetric shift in band edge energy also appears in H-TiClI and T-BiClTe, as a result of the orbital character of the band edge wave function\cite{riis2019classifying}.

\section{Environmental screening on quasiparticle band alignment}

Generally, environmental screening is expected to decrease the electronic band gap of isolated 2D materials, and thus affect the band alignment accordingly\cite{ugeda2014giant,qiu2017environmental,bradley2015probing}. To investigate the effect of environmental screening on the quasiparticle band alignment, we perform the GW calculation on monolayer MoSi$_2$N$_4$ with WSi$_2$N$_4$ substrate and compare it to the suspended one. Here, we only use the AA-stacked MoSi$_2$N$_4$/WSi$_2$N$_4$ heterostructure as an effective model to study the environmental screening effect of WSi$_2$N$_4$ on MoSi$_2$N$_2$, and other stacking modes will be discussed in section VI.

\begin{figure*}
\centering
\includegraphics[scale=0.47]{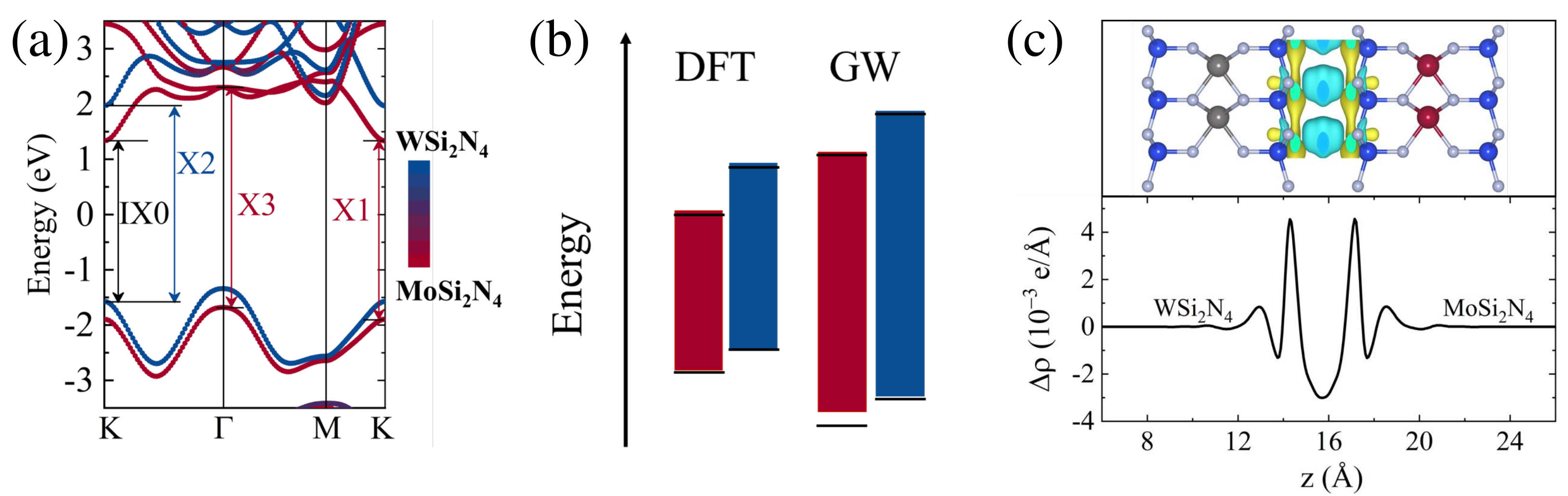}
\caption{The quasiparticle band structure of AA-stacked MoSi$_2$N$_4$/WSi$_2$N$_4$ heterostructure. IX$_0$, X$_1$, X$_2$, and X$_3$ label the intersubband transitions at K and $\Gamma$ that give rise to the corresponding excitonic or interband-transition peaks in Fig. 5. (b) Band alignments of monolayer MoSi$_2$N$_4$ (WSi$_2$N$_4$) in AA-stacked MoSi$_2$N$_4$/WSi$_2$N$_4$ structure calculated by PBE and GW calculations, where the band alignments of suspended monolayer (black lines) are listed for comparison. The red and blue rectangle represent the band alignments of MoSi$_2$N$_4$ and WSi$_2$N$_4$, respectively. (c) Plane-averaged charge density difference $\Delta \rho$(x) of MoSi$_2$N$_4$/WSi$_2$N$_4$ heterostructure. The top panel is a three-dimensional charge density difference, and the isosurface value is 2$\times$10$^{-5}$ e/Å$^3$. The yellow and blue areas represent electron accumulation and depletion, respectively.}
\end{figure*}

For AA-stacked MoSi$_2$N$_4$/WSi$_2$N$_4$, the Mo (Si, N) atoms in one layer fully overlap with the W (Si, N) atoms in the WSi$_2$N$_4$ layer. The lattice mismatch is nearly zero, and the optimized layer distance between Mo and W atom is 10.492 Å, much larger than that (6.23–6.54 Å) of bilayer TMDCs, such as MoS$_2$ and WS$_2$\cite{gao2017interlayer,he2014stacking}. This suggests that, compared with bilayer TMDCs, the interlayer vdW interaction in MoSi$_2$N$_4$/WSi$_2$N$_4$ heterostructure is much weaker, which is owing to the two passivated SiN$_2$ pyramid layers and has been observed in bilayer MoSi$_2$N$_4$\cite{zhong2021strain}. Such weak interlayer interaction leads to no band hybridization between two layers in the band edge of MoSi$_2$N$_4$/WSi$_2$N$_4$ heterostructure, as shown in Fig. 3a. This is totally different from bilayer or bulk TMDC, where the hybridization of chalcogen $p$-states at the $\Gamma$ point results in the crossover from direct band gaps in monolayers to indirect band gaps in multilayers\cite{zhang2014direct}. In detail, the VBM and CBM of MoSi$_2$N$_4$/WSi$_2$N$_4$ heterostructure are contributed by WSi$_2$N$_4$ and MoSi$_2$N$_4$ layer, respectively, demonstrating the type-II heterostructure, in agreement with the band alignment analysis in section IV. The VBM is dominated by W $d_{z^2}$ and $d_{x^2-y^2}$ orbitals, while CBM mainly consists of Mo $d_{z^2}$ orbitals, accompanied by N $p$ orbitals. On the whole, it is clearly shown that the indirect bandgap feature and band dispersion of monolayer MoSi$_2$N$_4$ is almost not affected by the WSi$_2$N$_4$ substrate, and vice versa. Therefore, the effect of wavefunction overlap may be ignored in MA$_2$Z$_4$ monolayers with substrates.

To study the environmental screening effect of WSi$_2$N$_4$ (MoSi$_2$N$_4$) in MoSi$_2$N$_4$/WSi$_2$N$_4$ structure, we plot the absolute band edge energies of monolayer MoSi$_2$N$_4$ (WSi$_2$N$_4$) in MoSi$_2$N$_4$/WSi$_2$N$_4$ structure in Fig. 3b. The values of VBM and CBM in MoSi$_2$N$_4$ are lower than those of WSi$_2$N$_4$, demonstrating type-II band alignment of MoSi$_2$N$_4$/WSi$_2$N$_4$ heterostructure, which is consistent with the projected band structure analysis in Fig. 3a. At the DFT level, the absolute VBM and CBM of monolayer MoSi$_2$N$_4$ (WSi$_2$N$_4$) in the contact system are very close to those of free-standing sample, with energy difference smaller than 0.05 eV. After including quasiparticle correction, compared with the suspended monolayer, the supported MoSi$_2$N$_4$ (WSi$_2$N$_4$) possesses almost the same CBM energy, while with higher VBM. The quasiparticle correction pushes up the VBM by 0.153 eV (0.051 eV) for supported MoS$_2$N4 (WSi$_2$N$_4$), leading to a 4.9$\%$ (1.7$\%$) reduction in quasiparticle band gap compared to free-standing monolayers. Comparing the band edge energies at DFT and GW levels, we can conclude that this environment-induced renormalization of the quasiparticle band gaps is a result of many-body effects on the screening and thus not apparent at the DFT level. On the other hand, the renormalization of the quasiparticle band gap of supported MoSi$_2$N$_4$ is the same as the bilayer MoSi$_2$N$_4$\cite{wu2022prediction}, confirming the reliability of our results. This quasiparticle band gap renormalization is very small in comparison with those of encapsulated monolayer black phosphorus (MoSe$_2$), whose quasiparticle band gap is reduced by 25$\%$ (11$\%$)\cite{ugeda2014giant,qiu2017environmental}. The small renormalization originates from the outer passivated A-Z bilayers in MA$_2$Z$_4$, protecting the band edge states from the dielectric screening. In this sense, our absolute band edge energy of isolated monolayer MA$_2$Z$_4$ in section IV is still instructive for realistic conditions with surrounded dielectric environment.

Apart from dielectric screening, charge transfer could also affect the band alignment of MoSi$_2$N$_4$/WSi$_2$N$_4$ heterostructure. In order to quantify this effect, we extract the plane-averaged charge density difference $\Delta \rho $(z) along the vertical direction (z axis) for MoSi$_2$N$_4$/WSi$_2$N$_4$ heterostructure in Fig. 3c. Here, $\Delta \rho $(z) is calculated by the charge density difference between the heterostructure and two noninteracting monolayers. For MoSi$_2$N$_4$/WSi$_2$N$_4$ heterostructure, the $\Delta \rho $(z) is on the order of 10$^{-3}$ e/Å, which is one order of magnitude smaller than that of typic type-II TMDC heterostructure like PtS$_2$/MoTe$_2$\cite{yin2022type}. This suggests small charge transfer between layers, consistent with the observed very weak interlayer vdW interaction in MoSi$_2$N$_4$/WSi$_2$N$_4$ heterostructure. On the whole, we can see clearly that changes in $\Delta \rho $(z) are almost symmetric on the two sides of the interface. This is very different from the asymmetric $\Delta \rho $(z) for other bilayers such as strained bilayer MoSi$_2$N$_4$ and TMDC\cite{yin2022type,zhong2021strain}. This symmetric charge difference suggests that the charge transfer is very small in MoSi$_2$N$_4$/WSi$_2$N$_4$ heterostructure, leading to the weak interlayer interaction in MA$_2$Z$_4$. It is the negligible charge transfer between the substrate and MA$_2$Z$_4$ that results in insignificant substrate effect for MA$_2$Z$_4$.

\section{Stacking-independent excitons in MoSi$_2$N$_4$/WSi$_2$N$_4$ heterostructure}

\begin{figure*}
\centering
\includegraphics[scale=0.51]{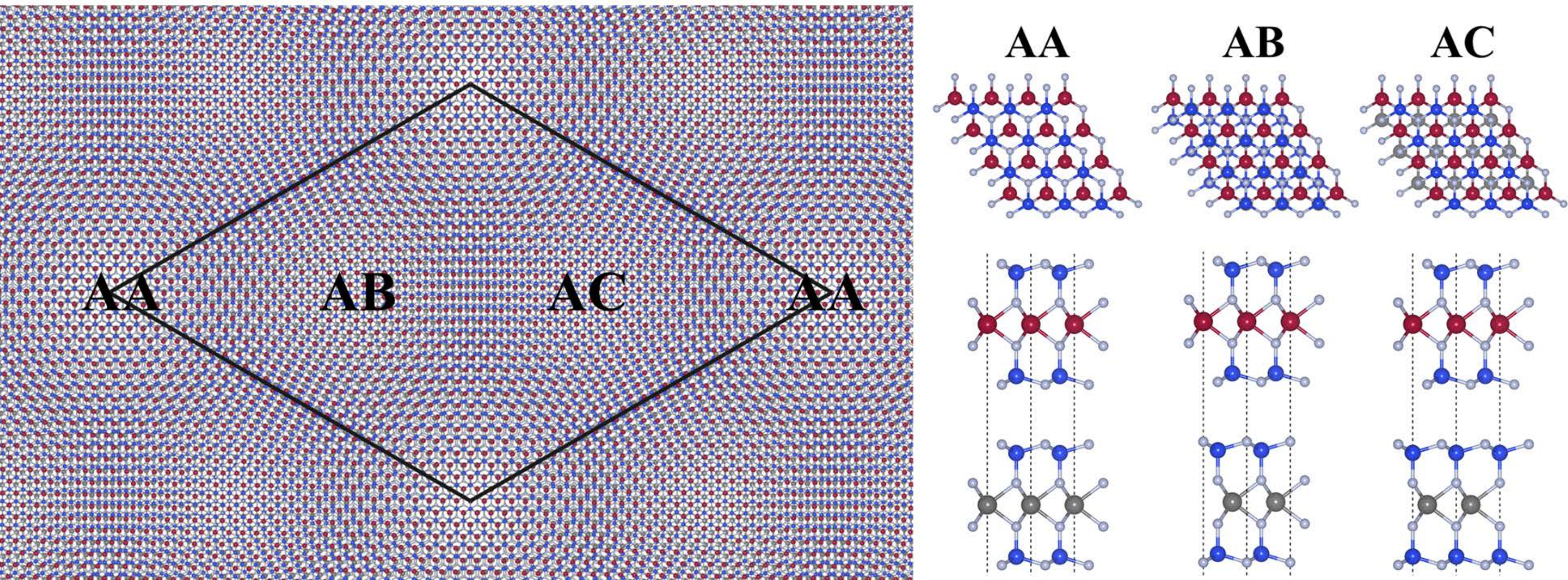}
\caption{The schematic plots of MoSi$_2$N$_4$/WSi$_2$N$_4$ heterostructure rotated from the AA stacking style with a small twist angle (moiré pattern). Three local stacking styles are identified and amplified with top and side views.}
\end{figure*}

In a real MoSi$_2$N$_4$/WSi$_2$N$_4$ heterostructure, two individual layers have different orientations, forming a twist angle and a moiré superlattice with a relatively large period. A small-angle moiré superlattice can be viewed as regions of high-symmetry stacking separated by domain walls\cite{ochoa2019moire}, in which the periodic modulation of local potential give rise to an exciton lattice centered on high-symmetry stacking region forms\cite{kennes2021moire,yu2017moire}. Hence, we will pay attention to specific stacking styles and calculate their local exciton properties. As shown in Fig. 4, we take the R type as an example to study the twist-angle dependent exciton properties of twisted MoSi$_2$N$_4$/WSi$_2$N$_4$ heterostructure. The R type represents a small twist angle rotated from the AA stacking style. In the R type of twisted bilayers, three local stacking styles can be identified in Fig. 4, which have been denoted as AA, AB, and AC, respectively. In the AA configuration, two monolayers are aligned, and all atoms of the same type are superimposed. Based on the AA configuration, the AB and AC configurations are obtained by shifting the bottom WSi$_2$N$_4$ layer along the long-diagonal of the unit cell by 1/3 and 2/3, respectively. The interlayer distances and relative energies of the three heterostructure structures are given in Table II. Among the three configurations, AA possess the largest interlayer distance (10.492 Å), and thus the weakest interlayer interaction. The largest interlayer distance is attributed to the repulsion arising from the N atoms superimposing in the two layers, leading to the highest relative energy. Meanwhile, AB and AC configurations share similar and lower interlayer distances and lower relative energies.

\begin{table*}
    \centering
    \caption{Interlayer distance $d$, relative energy $\Delta E$, band gap, and energy and optical oscillator strength of the excitons for the three different MoSi$_2$N$_4$/WSi$_2$N$_4$ heterostructures.}
    \begin{tabular}{ c c c c c c c c c c c c c c c c}
    \hline
    \hline
         &   &   &   &    &   &      & X$_1$ &   & IX$_0$ &   & X$_2$ &  & X$_3$ &    \\
    & $d$  & $\Delta E$  & $E_g^{DFT}$  &  $E_g^{GW}$  & $E_{gKK}^{DFT}$  & $E_{gKK}^{GW}$  & Energy & Osc. Str. & Energy & Osc. Str. & Energy & Osc. Str. & Energy & Osc. Str. \\
    &  (Å) &  (meV) &  (eV) &   (eV) &  (eV) &  (eV) & (eV) & (a.u.)   & (eV)   & (a.u.) & (eV) & (a.u.) & (eV) & (a.u.)\\
    \hline
    AA & 10.492 & 46 & 1.558 & 2.771 & 1.802 & 2.923 & 2.327 & 635 & 2.451 & 1.8$\times$10$^{-4}$ &	2.757 & 593 & 2.864 & 1672 \\
    
    AB & 10.144 & 0 & 1.454 & 2.652 & 1.706 & 2.891 & 2.298 & 652 & 2.505 & 3.1$\times$10$^{-4}$ & 2.767 & 690 & 2.866 & 1014  \\
    
    AC & 10.144 & 0 & 1.643 & 2.849 & 1.902 & 3.057  & 2.326 & 662 & 2.480 & 0.7$\times$10$^{-4}$ & 2.720 & 562 & 2.847 & 1199 \\
    \hline
    \hline
    \end{tabular}
    \label{tab:my_label}
\end{table*}

We first compare the quasiparticle band structure of the three stacking styles. Similar to AA configuration in Fig. 3a, a typical type-II quasiparticle band alignment is obtained in AB and AC modes. We focus on the band gap at K point, because the vertical inter-band transitions and excitons around these points are likely responsible for optical spectra observed in the MoSi$_2$N$_4$/WSi$_2$N$_4$ moiré heterostructure. Table II summarizes the GW-calculated quasiparticle band gaps at the K point, which vary with the local stacking styles. For the R stacking styles, the energy variation is observable: AB mode has the smallest band gap of 2.891 eV, while AC mode possess the largest band gap of 3.057 eV, showing a 166-meV variation of the quasiparticle band gap. This large quasiparticle band gap variation is larger than that (100-meV) of R-type MoSe$_2$/WSe$_2$ twisted bilayers\cite{lu2019modulated}.

\begin{figure}
\centering
\includegraphics[scale=0.26]{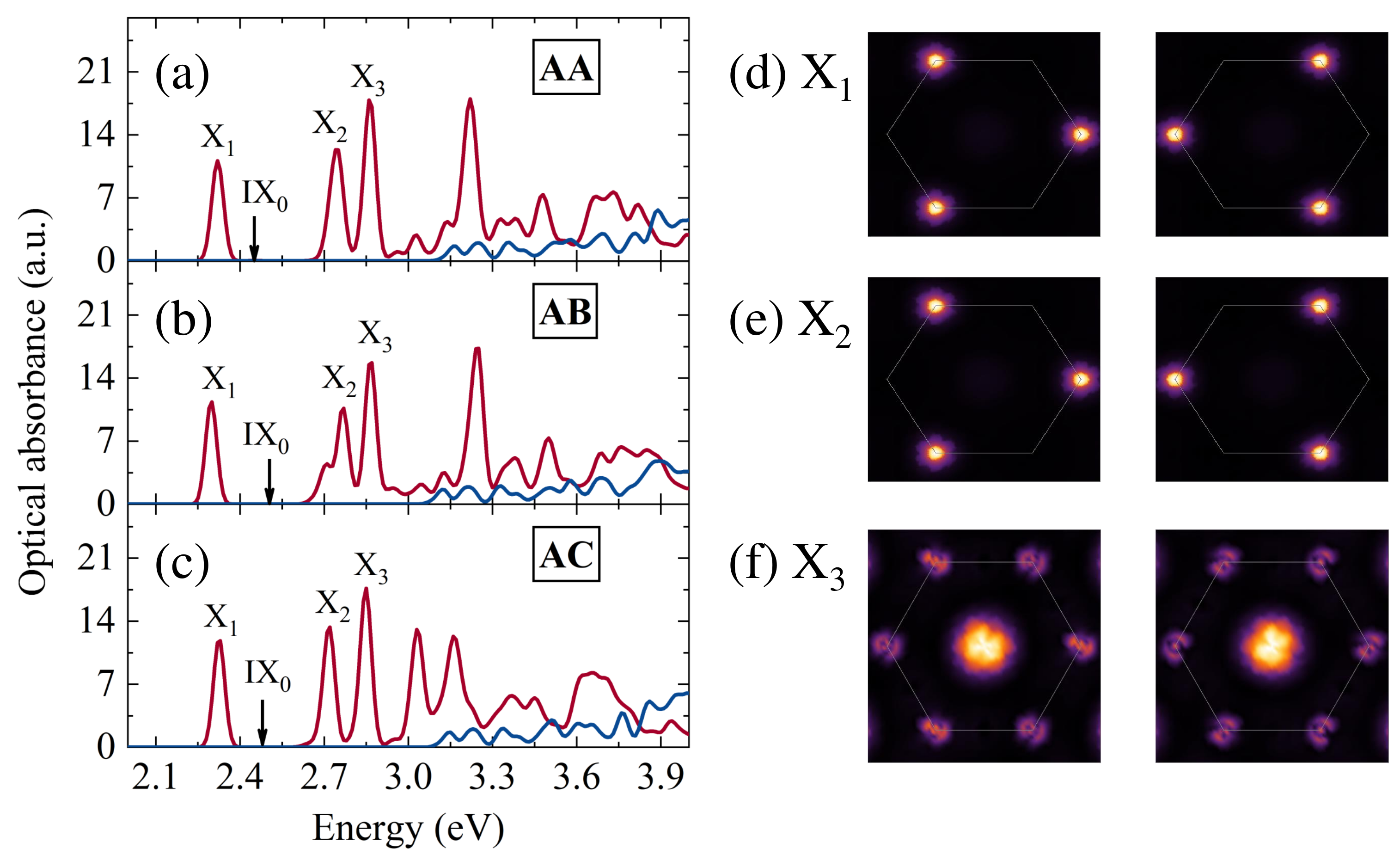}
\caption{(a)-(c)Optical absorption spectra of three identified local stacking styles in a MoSi$_2$N$_4$/WSi$_2$N$_4$ heterostructure  with (red) and without e-h (blue) interactions. The energy of the interlayer exciton (IX$_0$) is marked by the black arrow. A 20-meV smearing to spectral widths is applied. (d)-(f)The reciprocal-space distribution of the charge density of bright exciton X$_1$, X$_2$, and X$_3$.}
\end{figure}

Figure 5 shows the optical absorption spectra of the three local stacking styles. Like many other 2D structures, enhanced excitonic effects are observed: after including e-h interactions, three excitonic peaks are formed below the quasiparticle band gap with significant e-h binding energies around a few hundred meV. Generally, there are two types of excitons, the intralayer (X$_1$, X$_2$, and X$_3$) and interlayer ones (IX$_0$). To elucidate these features, we break down each exciton state into its component transitions. The exciton wave function can be written as a linear combination of electron-hole pairs
\begin{equation*}
    \Psi_\lambda(r_e{,}r_h{)}= \sum_{\upsilon c\mathbf{k}}A_{\upsilon c\mathbf{k}}^\lambda\psi_{c\mathbf{k}}(r_e{)}\psi_{\upsilon \mathbf{k}}^\ast(r_h{)}
\end{equation*}
where $\psi_{ck}(r_e{)}$ ($\psi_{vk}^\ast(r_h{)}$) is the quasi-particle electron (hole) wavefunctions; $\lambda $ indexes the exciton state; $v$ and $c$ index the occupied and unoccupied bands, respectively; and $A_{\upsilon ck}^\lambda$ is the electron-hole amplitude.

We now discuss the interlayer exciton (IX$_0$), marked by the black arrows in Fig. 5. For AA stacking mode, the exciton IX$_0$ comes primarily from transitions between the VBM and the CBM (Table III), which make up 99$\%$ of the band-to-band transitions composing the exciton and confirm the interlayer feature. The interlayer exciton IX$_0$ is located at 2.451 eV, and the quasiparticle direct band gap is 2.923 eV, resulting in an e-h binding energy of 472 meV. This is similar to the binding energy (410 meV) of MoSe$_2$/WSe$_2$ bilayers\cite{lu2019modulated}. On the other hand, the optical oscillator strength of the interlayer exciton is very small (on the order of 10$^{-6}$ times that of the intralayer exciton), as a result of the complete separation of electron and hole wave functions. This dark interlayer exciton is very different from the stacking-dependent interlayer excitons in TMDCs, with dipole oscillator strength reaching the same order as those of intralayer excitons. Moreover, unlike what is normally expected for type-II heterostructures, the interlayer exciton is not the lowest-energy one, owing to the decreased coulomb interaction in the thick MoSi$_2$N$_4$/WSi$_2$N$_4$ heterostructure. Finally, the energy and brightness of the interlayer exciton are nearly independent of the stacking style in Table II. For example, the interlayer exciton located at 2.451 eV in AA mode, 2.505 in AB mode, and 2.480 eV in AC mode, and the corresponding exciton binding energies are 472, 386, and 577 meV, respectively. The order of the exciton binding energy is consistent with the direct quasiparticle band gap at K point, following the linear scaling law between the exciton binding energies and quasiparticle band gaps observed in 2D materials\cite{choi2015linear}. This suggests that the moiré superlattice in twisted bilayer MoSi$_2$N$_4$/WSi$_2$N$_4$ does not modulate the interlayer exciton properties.

There are three main bright exciton peaks X$_1$, X$_2$, and X$_3$ in the optical spectra below the quasiparticle direct gap. The exciton wavefunction shows that the first peak X$_1$ at lowest-energy (2.327 eV) arises from two degenerate excitonic states, and comes from transitions between VBM-1 and CBM (Table III) within MoSi$_2$N$_4$ monolayer, indicating the intralayer character. Figure 5 shows the $k$-resolved e-h pair amplitudes for these two excitonic states, which are dominated by the e-h pairs near the minimum direct gap at K and K$^{'}$ points, similar to that in monolayer MoS$_2$\cite{qiu2013optical}. While for the second peak X$_2$ around 2.757 eV, it comes from transitions between VBM and CBM+1, i.e., coming from WSi$_2$N$_4$ layer. The exciton X$_2$ originates from the direct transition at K-point in reciprocal-space exciton wave function. At energies above the second peak, the third peak X$_3$ is due to the direct transition at $\Gamma$ point partially coinciding with other direct transition at K point, mainly dominated by MoSi$_2$N$_4$ layer. Different from the negligible dipole oscillator strength in dark interlayer exciton, the bright intralayer excitons have large dipole oscillator strength, because of the significant overlap of their electron and hole wave functions. Strikingly, the optical oscillator strengths of these three intralayer excitons (listed in Table II) are robust to the stacking style, having the same order of 10$^2$ for excitons X$_1$ and X$_2$, and the order of 10$^3$ for exciton X$_3$. This is very different from the excitons in TDMC moiré supercells, with the optical dipole oscillator strength modulated by a few orders of magnitude. These results are valuable for designing material platforms of Moiré excitons and exciton condensation.

\begin{table}
    \centering
    \caption{Dominant band-to-band transition and its weight $\sum_k|A_{vc\mathbf{k}}|^2$ of excitons in MoSi$_2$N$_4$/WSi$_2$N$_4$ heterostructures.}
    \begin{tabular}{c c c c c }
    \hline
    \hline
  State  & Conduction band & Valence band  & $\sum_k|A_{vc\mathbf{k}}|^2$ \\
    \hline
    IX$_0$ & CBM & VBM & 0.992 \\
    X$_1$ & CBM & VBM-1 & 0.989 \\
    X$_2$ & CBM+1 & VBM & 0.997 \\
    X$_3$ & CBM & VBM-1 & 0.654 \\
    \hline
    \hline
    \end{tabular}
    \label{tab:my_label}
\end{table}

\section{Conclusion}
In conclusion, we systematically study the quasiparticle band alignment and exciton in monolayer MA$_2$Z$_4$ (M = Mo, W, Ti; A= Si, Ge; Z = N, P, As). Compared with the results from DFT, monolayer MA$_2$Z$_4$ possess substantially larger quasiparticle band gaps, and the different absolute quasiparticle band alignments accordingly. On the other hand, the band-gap-center model works very well for obtaining the absolute band edge energy of monolayer MA$_2$Z$_4$, and the qualitative types of band alignments for these materials have not changed from DFT and GW. Based on the obtained band alignments, we design type-II heterostructure MoSi$_2$N$_4$/WSi$_2$N$_4$ without lattice mismatch and use this mode to discuss the effect of environmental screening and moiré pattern on quasiparticle band alignment and excitons. It is found that the absolute quasiparticle band alignments are robust to the environmental screening, owing to protecting effect of the capped A-Z sublayers in MA$_2$Z$_4$. Finally, we study interlayer and intralayer excitons in MoSi$_2$N$_4$/WSi$_2$N$_4$ bilayers. Our calculations reveal that the dipole oscillator strength and energy of intralayer excitons are almost independent of the local stacking configuration, suggesting the negligible effect of moiré pattern modulation. The absolute band edge energies and band offsets obtained in this work are important for designing heterojunction devices based on monolayer MA$_2$Z$_4$.

\section*{Acknowledgement}
This work is supported by the National Natural Science Foundation of China (Grant Nos. 12104421 and 11947218). SG is supported as part of the Institute for Quantum Matter, an Energy Frontier Research Center funded by the U.S. Department of Energy, Office of Science, Basic Energy Sciences under Award No. DE-SC0019331. Numerical calculations presented in this paper have been performed on a supercomputing system in the Supercomputing Center of Wuhan University.


\bibliography{apssamp}

\end{document}